\documentclass[journal,onecolumn]{IEEEtran}
\ifCLASSINFOpdf
\else
\fi

\usepackage{graphics} 
\usepackage{graphicx}
\usepackage{float}
\usepackage{placeins}
\usepackage{amssymb}
\graphicspath{{./imgs/}}
\usepackage{caption}
\usepackage[dvipsnames]{xcolor}
\usepackage{url}
\begin{document}
%
\title{Dense Passage Retrieval in Conversational Search}

\author{Ahmed~H.~Salamah\footnotemark*, Pierre~McWhannel\footnotemark*,\and~Nicole~Yan\footnotemark*
         \thanks{\footnotemark* Equal contributions.}}



\markboth{STAT946: Deep learning, December~2020}%
{Ahmed H. Salamah \MakeLowercase{\textit{et al.}}: Bare Demo of IEEEtran.cls for IEEE Journals}

\maketitle

\begin{abstract}
Information retrieval systems have traditionally relied on exact term match methods such as BM25 for first-stage retrieval. However, recent advancements in neural network-based techniques have introduced a new method called dense retrieval. This approach uses a dual-encoder to create contextual embeddings that can be indexed and clustered efficiently at run-time, resulting in improved retrieval performance in Open-domain Question Answering systems.
In this paper, we apply the dense retrieval technique to conversational search by conducting experiments on the CAsT benchmark dataset. We also propose an end-to-end conversational search system called GPT2QR+DPR, which incorporates various query reformulation strategies to improve retrieval accuracy. Our findings indicate that dense retrieval outperforms BM25 even without extensive fine-tuning.
Our work contributes to the growing body of research on neural-based retrieval methods in conversational search, and highlights the potential of dense retrieval in improving retrieval accuracy in conversational search systems.


\end{abstract}


\IEEEpeerreviewmaketitle

\section{Introduction} \label{sec:Intro}
Conversational search allows users to speak into search engines and express their needs in a conversational manner. Conversational queries are typically multi-turn, dependent, and expressed informally with word omission, which makes identifying information needs hard. Another challenge is to efficiently retrieve relevant passages from a large corpus. This problem is studied in Open-Domain Question Answering (QA). A common method is a two-stage framework: a \emph{retriever}, usually BM25, which selects a subset of passages by counting weighted overlapping words, and a neural \emph{reader} which selects the answer(s) among the subset. One limitation of this is that term-frequency based first stage retrievers might fail to capture semantic relationships. An alternative way is to use pre-trained language models such as BERT to encode passages in a semantic space so that similar queries and passages are close to each other. Dense vectors can be efficiently indexed and clustered, which speeds up the retrieval process. This approach is called dense retrieval, and has proven to yield better QA system performance \cite{karpukhin2020dense}. Inspired by this, we employed dense passage retrieval (DPR)\cite{karpukhin2020dense} in conversational search. We worked on the 2019 \textbf{C}onversational \textbf{As}sistance \textbf{T}rack (CAsT) \cite{dalton2020trec}, which is a conversational search benchmark at \textbf{T}ext \textbf{RE}trieval \textbf{C}onference (TREC).

The task of CAsT is to retrieve relevant passages given a sequence of queries in a dialogue format. CAsT organizers provided 108 labeled conversational utterances in 13 topics, and the passage collection was built from MSMARCO Passage ranking \cite{bajaj2018ms} and Wikipedia TREC CAR \cite{CAR2018}. Previous CAsT participants usually use BM25-BERT cascade systems. Our method only uses a dense retriever. Due to the limited amount of labeled utterances, it's infeasible to train a model on CAsT data directly. We therefore adopted transfer learning and planned to train the DPR model on MSMARCO data, which has 530k queries with relevant passages. It took 9+ hours to complete one epoch using 2 GPUs on ComputeCanada. Due to resource limitation and time constraints, we decided to use the DPR trained on five QA datasets, and fine-tune it on CAsT.  

\subsection{Motivation}
Downstream NLP tasks benefit a lot from pre-trained language models. However, first stage information retrieval still relies on bag-of-words models such as BM25, which ranks documents based on term-frequency and could fail to identify semantic relationships. Dense retrieval is a new technique and works well on QA systems. We therefore want to adopt this state-of-the-art method in conversational search and evaluate its performance. 

\subsection{Overview}
We break our research into two sub-tasks: query reformulation and dense retrieval. Query reformulation makes raw utterances more informative by integrating previous context information. We tried three methods: (1) AllenNLP coreference resolution, (2) Attention-based Bi-GRU encoder-decoder, and (3) a fine-tuned GPT2. By using dual-enocoders, passage embeddings can be pre-computed offline, and only query embeddings are computed at run-time, followed by a dot-product and Maximum Inner Product Search (MIPS) \cite{guo2015quantization} to identify candidate passages. This technique is employed in Dense Passage Retrieval (DPR) \cite{karpukhin2020dense}.

DPR has been shown to outperform Lucene-BM25 on top-20 passage retrieval accuracy on multiple QA benchmarks NQ, TriviaQA, etc\cite{Kwiatkowski2019NaturalQA, joshi2017triviaqa, berant-etal-2013-semantic, Baudis2015ModelingOT}. It is trained with in-batch negative sampling and one additional hard negative. In-batch negative sampling means using positive passages for other queries in the same mini-batch as negatives, and hard negatives refer to the irrelevant passages that are highly ranked by BM25. We initially planned to train DPR on MSMARCO using hard negatives, and we selected hard negatives by verifying if the results returned by BM25 were labeled negative in MSMARCO. However, after investigation, we found MSMARCO assessors only judged a subset of the passage collection. Unjudged passages and irrelevant passages are both labeled negative. In other words, negative passages consist of true negatives and false negatives. Furthermore, researchers have found 70\% of the negative passages are relevant \cite{ding2020rocketqa}, which made it impossible to select hard negatives from MSMARCO. Therefore, we only used in-batch negative sampling. 

In our work we experiment with two main query reformulation strategies and apply DPR on the 2019 TREC CAsT task. Additionally, we provide analysis of the DPR results on the CAsT evaluation data. Our contributions are primarily as follows:

\begin{enumerate}
  \item Designed and evaluated different query reformulation strategies.
  \item Applied and evaluated DPR technique in the conversational search benchmark CAsT and compared with other participants.
  \item Built an end-to-end conversational search system (GPT2QR+DPR) for the CAsT task.
\end{enumerate}


\begin{table}[H]
    \caption{CAsT Evaluation Example of Topic 33  \cite{dalton2020trec}
    \label{tab:utterances}}
    \centering
    \begin{minipage}{.4\linewidth}
      \centering
        \begin{tabular*}{\linewidth}{@{\extracolsep{\fill}}l|l}
        \hline
        \textbf{Turn} & \textbf{Raw Conversation Utterances} \\ \hline
          & \\
        1 &	Tell me about the Neverending Story film. \\
        2 &	What is it about? \\
        3 &	How was it received? \\
        4 &	Did it win any awards? \\
        5 &	Was it a book first? \\
        6 &	Who was the author and when what it published? \\
          & \\
          & \\
        7 &	What are the main themes? \\
         & \\
        8 &	Who are the main characters? \\
          & \\
        \hline
        \end{tabular*}
    \end{minipage}%
    \hspace{1cm}
    \begin{minipage}{.4\linewidth}
      \centering
        \begin{tabular*}{\linewidth}{@{\extracolsep{\fill}}l|l}    
        \hline
        \textbf{Turn} & \textbf{Utterances Rewritten by GPT2QR} \\ \hline
         & \textbf{Handle co-ref correctly} \\
        1 &	Tell me about the Neverending Story film. \\
        2 &	What is the Neverending Story film about ? \\
        3 &	How was the Neverending Story film received ? \\
        4 &	Did the Neverending Story film win any awards ? \\
        5 &	Was the Neverendering Story film a book first ? \\
        6 &	Who was the author and when what the Neverendering ? \\
          & Story film published ? \\
          & \textbf{Handle omission correctly} \\
        7 &	What are the main themes of the Neverendering \\
          & Story film ? \\
        8 &	Who are the main characters in the Neverendering \\
          & Story film ? \\
        \hline
        \end{tabular*}
    \end{minipage} 
\end{table}

\section{Datasets}

TREC CAsT is a multi-turn conversational search task as it contains a large-scale reusable test collection for sequences of conversational queries. The queries of CAsT exhibit three distinct properties (1) a sequential dialog history, (2) co-reference to previous utterances/queries, and (3) word omission within a query. These challenges necessitate a query reformulation for effective retrieval beyond a traditional IR system as tracking and understanding of the history is necessary. Additionally, in CAsT the utterances do not need information from the retrieval for reformulation though topic shift still exists. The CAsT evaluation set consists of 20 topics and 173 turns the results are based on this set.

The large scale collection to meet the queries of TREC CAsT are comprised of two collections MSMARCO, TREC CAR, and the \textbf{Wa}shington \textbf{Po}st Collection (WaPo); though WaPo was removed from the creators evaluation due to an ambiguous document id error. MSMARCO consists of 8,841,823 passages and additionally has its own 592,034 of its own open-domain questions which have at least one associated relevant passage.  TREC CAR is a larger collection than MSMARCO consisting of approximately 30 million unique paragraphs.
MSMARCO provides a larger dataset to further fine-tune DPR upon; however, the pool of the top-1000 retrieved by BM-25 was not fully judged; researchers \cite{ding2020rocketqa} found it to contain 70\% hard negatives. This is important as selecting negatives (Gold/Hard negatives) from this pool improved DPR's performance \cite{karpukhin2020dense}, this limits training to in-batch negatives.

Finally, a new dataset \textbf{Q}uestion \textbf{Re}writing in \textbf{C}onversational \textbf{C}ontext (QReCC) \cite{anantha2020opendomain} containing approximately 14,000 conversations and 81,000 question-answer pairs. This dataset is constructed from a combination of QuAC\cite{choi2018quac}, NQ and CAsT. This dataset has conversational questions re-written allowing an opportunity to train a model to learn the three distinct properties of CAsT's queries. Then with transfer learning a model trained on QReCC can be leveraged to perform query reformulation on the TREC CAsT dataset.


\subsection{MSMARCO}
The MAchine Reading COomprehension dataset named (MSMARCO) where MS comes from MicroSoft is a large scale dataset. This pieces of this dataset of use to our research our the 8,841,823 passages which were extracted from 3,563,535 web documents retrieved by Bing. This dataset consists of a total of 592,034 queries with at least one associated relevant passage. Additionally, they provide the top 1000 passages retrieved by a BM25 for each query we wished to utilize this to select hard/gold negatives, these are negatives with high lexical overlap but are negative. However, these subsets of passages often contain false negatives as judges did look through all 1000. The researchers of RocketQA even found that 70\% of the passages in the top 1000 are either positive or highly relevant. This was the primary motivators for us to utilize ANCE in our training.

\section{Frameworks}
To test the dense representation in DPR, we were curious to understand if it is still robust enough to deal with another task that the model did not train on or not. In Table \ref{tab:Automatic}, we show our inference results for the CAsT dataset to reach $6^{th}$ rank with just inferring on the training model of the DPR. This model was trained on Multi-sets that is mentioned  in section \ref{sec:Intro}. This initial results motivated us to one step forward to make our analysis and comparison between DPR and ANEC frameworks.

\subsection{Query Reformulation and Dense Retrieval}

Two query reformulations (QR) were assessed AllenNLP's spanBERT and a transformer decoder \cite{radford2019language} pre-trained on QReCC. Additionally, we build a Bi-GRU with attention and GloVE embeddings \cite{pennington2014glove} trained on QReCC; however, this model produced nonsensical QR. Our transformer decoder model used the teacher forcing approach, where it predicts the following token in the desired output sequence and then uses the true output token in the next step. The input sequence was pre-appended with $<|startoftext|>$, previous questions were separated by $<|sep|>$, the query to reformulate appended with $<|go|>$, and then the true rewritten query follows with $<|endoftext|>$ token appended and then padded to a maximum sequence length of 210. A mask was used on the cross-entropy loss to solely focus on the re-written tokens. This model was instantiated with the GPT-2-small \cite{radford2019language} weights as the transformer decoders, thereby consisting of 12 layers and trained on QReCC for 3 epochs. Table \ref{tab:utterances} shows an example of conversational utterances and reformulations by our GPT2QR. As can been observed GPT2QR succeeds in co-reference resolution and word omission.

\subsection{Dense Retrieval}
Dense Passage Retrieval (DPR) is a first stage retriever having an architecture of dual-encoders where each encoder is an instantiation of BERT \cite{devlin2019bert}. These encoders encode the query and passage into a dense vector space where relevant pairs are placed closer together than irrelevant ones. DPR uses the following similarity function to measure the relevance of a query $q$ and passage $p$:

$$sim(\vec{e}_q,\vec{e}_p) = E_{Q}(\vec{e}_q)^{T} E_{P}(\vec{e}_p) \in \mathbb{R}$$

The model is then trained using the negative log likelihood as seen in the following equation:
$$L(\vec{e}_{q,i},\vec{e}_{p,i}^{+},\vec{e}_{p,i,1}^{-},...,\vec{e}_{p,i,n}^{-}) = -log \frac{e^{sim(\vec{e}_{q,i},\vec{e}_{p,i}^{+})}}{e^{sim(\vec{e}_{q,i},\vec{e}_{p,i}^{+})} + \sum^{n}_{j=1} e^{sim(\vec{e}_{q,i},\vec{e}_{q,i,j}^{-})}}$$

As seen the conditional probability $Pr(\vec{e}_{p,i}^{+} | \vec{e}_{q,i})$ is estimated using a sampled SoftMax for practicality. Additionally, DPR uses in-batch negatives for memory efficiency and one additional gold or hard negative that comes from a negative within the top-1000 retrieved passages from a BM25. Additionally, DPR can be indexes efficiently since the dense embeddings are in a lower dimensional continuous space. As previously mentioned in the introduction section we were unable to use hard-negatives due to lack of annotation, in the case of MSMARCO 70\% \cite{ding2020rocketqa} of the negative candidates in the top-1000 BM25 pool are false negatives. Therefore we used in-batch negatives for fine-tuning DPR and used DPR's \cite{karpukhin2020dense} model weights to instantiate our base DPR model.
To resolve this issue Approximate nearest neighbor Constrastive Estimation (ANCE) \cite{xiong2020approximate} is a computationally intensive technique for training a dense retriever. This technique asynchronously creates embeddings for the entire corpus from a checkpoint during training. Next the ANN index is updated so that negatives which are closest to the queries can be utilized for negative samples. As these samples are more informative during training and improve training convergence by creating large gradient norms. This technique looks feasible to amend our hard negative issue; however, was too resource extensive for this research endeavour. 

\subsection{GPT2QR+DPR: Conversational information seeking end-to-end model}
\begin{figure}[H]
    \centering
    \includegraphics[width=0.8\textwidth]{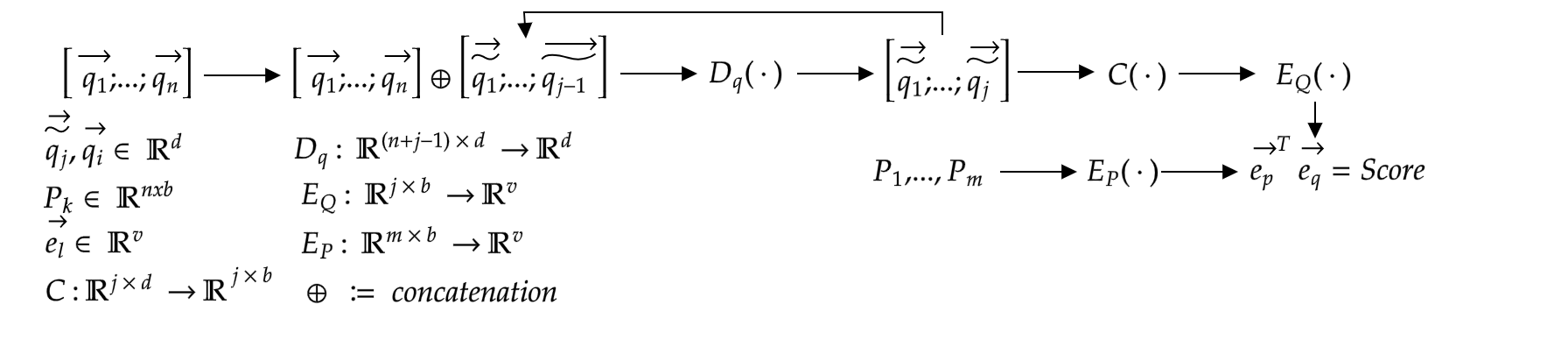}
    \caption{Complete model architecture of GPT2QR+DPR}
    \label{fig:model}
\end{figure}

Our end-to-end system at inference time can be seen in figure \ref{fig:model}. The tokenized queries $\vec{q_i},i=1,..,n$ are passed to the transformer decoder which then generates a token to concatenate to the input sequence. This auto-regressive process is repeated until the $<|endoftext|>$ token is generated then the GPT2 tokenizer decodes the tokens before being passing them to DPR model's query encoder $E_Q$. Offline the embeddings $\vec{e}_{p_i}, i=1,...,m$ are computed then at run time they are indexed. Finally, the dense embeddings are used to calculate a similarity score using a dot-product between the passage embeddings which were pre-computed offline. 

\section{Results and Analysis}
In this section, we present the experimental setup and show some challenges we had through setting up our environment. Then, we are going to do some analysis and benchmark our results relative to other submissions shown in \cite{dalton2020trec}.

The DPR base model is fine-tuned by exploiting CAsT training data. The model is fine tuned on the CAsT data \cite{dalton2020trec} which only provides a total of 108 conversational user utterances in 13 topics with relevance judgments. As the price of adopting two encoders is to use more data which might make their accuracy drastically varies across different hyperparameter configurations. In our experiments, we fine-tuned the DPR based on the way each method reformulates the queries from ambiguous state that is hard to interpret them without context. Although, the computational resources were a major drawback in our experiments. We tried to use Google Colabs (premium) and ComputeCanda with all clusters (Cedar, Graham, Beluga and Niagara) to train 10 epoch using 2 GPUs on MSMARCO. We were only able to train for 1 epoch for more than 9 hours on ComputeCanada with a small batch size, while in the original benchmark they have used $8\times32$GB GPUs \cite{karpukhin2020dense} for 25 epochs. After many trials, we did not manage to train the base-line encoder on MSMARCO. We managed to generate the passage embeddings by using dense encoder base model that is trained on multiset data and bert-based-uncased \footnote{\url{https://dl.fbaipublicfiles.com/dpr/checkpoint/retriver/multiset/hf\_bert\_base.cp}} used in the DPR paper. We have split the operation into 70 shards that took a total of 170 hours on ComputeCanada using 1 GPU, which limited us to a single fine-tuning as more variations would require repeating this process. Our code is available online on GitHub \footnote{\url{https://github.com/AhmedHussKhalifa/Dense\_Passage\_Retrieval\_in\_Conversational\_Search}}. 


The CAsT dataset contains \textbf{raw utterances} which require reformulation as shown in Table \ref{tab:utterances}. Also, it contains \textbf{manually} rewritten utterances containing all information needed from prior rounds as a single query. In our experiments, we considered manual runs as an empirical bound of human performance. Another baseline used in the literature and our experiments to resolve the raw utterances is \textbf{AllenNLP}. Additionally, in our analysis we considered pre-appending the $1^{st}$ utterance $u_{0}$ and the previous query $u_{i-1}$ for each reformulated query from \textbf{AllenNLP} and \textbf{GPT2QR}. We believe that including $u_{i-1}$ for each $u_{i}$ query will solve the subtopic shifting and convert the ad-hoc search session to better include the context-dependent information. 

\begin{table}[H]
    \centering
    \begin{minipage}{.45\linewidth}
      \centering
        \begin{tabular}{|l|c|c|c|}
        \hline
        \textbf{Run}  & \textbf{MAP} & \textbf{MRR} & \textbf{NDCG@3}\\
        \hline
        Raw+DPR & 0.089 & 0.308 & 0.145 \\
        AllenNLP+DPR & 0.109 & 0.344  & 0.168  \\
        AllenNLP+DPR$+u_{i-1}$ & 0.121 & 0.370 & 0.192 \\
        Raw+DPR$+u_{0}$ & 0.126 & 0.385 & 0.194 \\
        GPT2QR+DPR$+u_{0}$ & 0.128 & 0.385 & 0.200 \\
        GPT2QR+DPR & 0.133 & 0.411 & 0.212 \\
        AllenNLP+DPR$+u_{0}$ & \textbf{0.15} & 0.424 & 0.221 \\
        GPT2QR+DPR$+u_{i-1}$ & 0.138 & \textbf{0.440} & \textbf{0.229} \\
        Manual+DPR & \textbf{0.163} & \textbf{0.503} & \textbf{0.263} \\
        \hline
        Raw+DPR(FT) &  0.070 & 0.281 & 0.145 \\
        AllenNLP+DPR(FT)$+u_{0}$ &  0.126 & 0.447 & 0.226 \\
        Manual+DPR(FT) &  0.142 & 0.461 & 0.243 \\
        \hline 
        \end{tabular}
        \caption{Retrieval Results on MS-MARCO. 
        \label{tab:MSMARCO}}
        
    \end{minipage}%
\end{table}

\begin{table}[H]
    \centering
    \begin{minipage}{.45\linewidth}
      \centering
        
        \begin{tabular}{|l|l|c|c|c|}
        \multicolumn{5}{c}{\textbf{Manual Runs}} \\
        \hline
        \textbf{Run} & \textbf{Group} & \textbf{MAP} & \textbf{MRR} & \textbf{NDCG@3}\\
        \hline
        UMASS\_DMN\_V2 & UMass & 0.082 & 0.300 & 0.100\\
        ict\_wrfml & ICTNET & 0.105 & 0.373 & 0.165\\
        UNH-trema-ecn & TREMA-UNH & 0.073 & 0.505 & 0.222\\
        unh-trema-relco & TREMA-UNH & 0.077 & 0.533 & 0.239\\
        UNH-trema-ent & TREMA-UNH & 0.076 & 0.534 & 0.242\\
        \textbf{DPR} & \textbf{Ours} & 0.113 & 0.518 & 0.242 \\
         topicturnsort & ADAPT-DCU & 0.136 & 0.555 & 0.259\\
        \multicolumn{1}{|c|}{.}  & \multicolumn{1}{c|}{$ \uparrow$ } & . & . & .\\
        \multicolumn{1}{|c|}{.}  & \multicolumn{1}{c|}{15 groups } & . & . & .\\
        \multicolumn{1}{|c|}{.}  & \multicolumn{1}{c|}{$\downarrow$} & . & . & .\\
        CFDA\_CLIP\_RUN6 & CFDA\_CLIP & 0.392 & 0.861 & 0.572\\
        humanbert & ATeam & 0.405 & 0.879 & 0.589\\
        \hline
        \multicolumn{5}{c}{\textbf{Automatic Runs}} \\


        \hline
         SMNgate & RALI & 0.030 & 0.072 & 0.008\\
        ECNUICA\_BERT & ECNU-ICA & 0.008 & 0.106 & 0.021\\
        mpi-d5\_union & mpi-inf-d5 & 0.098 & 0.274 & 0.078\\
        MPmlp & RALI & 0.054 & 0.285 & 0.090\\
        SMNmlp &RALI & 0.060 & 0.244 & 0.090\\
        UMASS\_DMN\_V1 & UMass & 0.077 & 0.298 & 0.108\\
        MPgate & RALI & 0.053 & 0.282 & 0.108\\
        \textit{indri\_ql\_baseline} & - & 0.139 & 0.328 & 0.152\\
        galago\_rel\_q & USI & 0.105 & 0.394 & 0.181\\
        galago\_rel\_1st & USI & 0.112 & 0.426 & 0.197\\
        \textbf{AllenNLP+DPR+$u_{0}$} & \textbf{Ours} & 0.106 & 0.424 & 0.200\\
        ECNUICA\_MIX & ECNU-ICA & 0.171 & 0.522 & 0.231\\
        \multicolumn{1}{|c|}{.} & \multicolumn{1}{c|}{$ \uparrow $} & . & . & .\\
        \multicolumn{1}{|c|}{.} & \multicolumn{1}{c|}{31 groups } & . & . & .\\
        \multicolumn{1}{|c|}{.} & \multicolumn{1}{c|}{$\downarrow$} & . & . & .\\
        CFDA\_CLIP\_RUN7 & CFDA\_CLIP & 0.267 & 0.715 & 0.436\\
        \hline
        \end{tabular}
        \caption{Manual and Automatic response retrieval results \cite{dalton2020trec}. \label{tab:Automatic}}
    \end{minipage} 
\end{table}



\subsection{Evaluation}

The main metrics used in our analysis is the top-k ranking metric NDCG@3 averaged on all turns using uniform weights and we will consider this metric for comparison for simplicity as used in \cite{dalton2020trec} as all tables are ordered ascendingly based on NDCG@3. we also calculated the mean average precision (MAP) and the Mean Reciprocal Rank (MRR) at depth of 1000. These metrics are all used in \cite{dalton2020trec} to evaluate submitted runs and we evaluated using a standard tool used by all submissions \footnote{\url{https://github.com/usnistgov/trec\_eval}}; higher values for all three metrics correlate to better system performance.

We initiated our analysis in table \ref{tab:MSMARCO} by making the ad-hoc search sessions that are collected from MSMARCO \footnote{\url{https://github.com/microsoft/MSMARCO-Conversational-Search}} and fine tune the DPR on the training set of the CAsT using the \textbf{AllenNLP+$u_0$} queries. The motivation of this table is to explore different reformulation methods and DPR with and without fine tuning and then choose the trial with the highest score to run in Table \ref{tab:Automatic}. In table \ref{tab:MSMARCO}, it is shown that the raw and manual runs are respectively the lower and upper bounds for both the base model (DPR), which validates that we can even use the DPR based model in this task. In Tables \ref{tab:Automatic}, it is found that the highest Automatic run, scored a NDCG@3 of 0.436, trivially lower than the highest manual run of 0.572 for the same group (CFDA\_CLIP). This observation is aligned with our results in table \ref{tab:MSMARCO}.

 The base model DPR in table \ref{tab:MSMARCO} on Raw queries showed a significant change after adding the $u_0$ as well as AllenNLP that even over preformed the raw+$u_0$ by 0.29 points, demonstrating that adding extra information can be valuable for retrieval. While without added information our baseline GPT2QR will significantly superpass the AllenNLP by 0.44 points. In case of including the previous utterance $u_{i-1}$ with query, the GPT2QR will also defeat the AllenNLP by 0.37 points and only 0.34 points behind the manual run. After fine tuning the DPR using the AllenNLP+$u_0$ we see an increase by 0.05 points suggesting the fine tuning makes a slight improvement.We were not able to fine tune on our best trial of GPT2QR+DPR+$u_{i-1}$ due to resources and time limitation but we expect that It will outrank the AllenNLP+$u_0$ as observed for the base DPR model. In Table \ref{tab:Automatic}, we only managed to run one trial using the DPR base model to achieve a certain rank on the leading board. We believe that we can do better if we had the required resources and enough time.

\section{Conclusion and Future Work}
Instead of applying BM25 + re-ranked we applied the dense passage retrieval in the conversational search task and found that it produced comparable results, where the shown results in table \ref{tab:MSMARCO} could be improved with query reformulation, and fine tuning. We observe that including $u_{i-1}$ for each $u_{i}$ query is more robust than including $u_{0}$ with GPT2QR as it seems to help in solving the subtopic shifting and will convert the ad-hoc search session to incorporate context-dependent information. In our baseline, we did not use the BERT-based reranking methods as other groups have used in table \ref{tab:Automatic} and we might use BERT to rerank in our future experiments. 

In this problem, the probability of generating $p_{gen}$ a token to select the next token is taken from the SoftMax applied to the inner product of the word embedding matrix and the output hidden vector from GPT2. We believe adding this to a copy distribution from tokens in the previous questions could improve the performance $p_{copy}$. We believe this since the words used to reformulate the query often appear in the dialog history. The selection of the next query would then be $w p_{gen} + (1-w)p_{copy}$ where $w$ is a weight adding the distributions. This is motivated by pointer-generator networks and their success with text summarization. \cite{see2017point}.

Our future work will consider the GPT2QR baseline to reformulate this problem. GPT2QR's current query reformulation uses $p_{gen}$ the probability to select the next token as taken from the SoftMax applied to the inner product of the word embedding matrix and the output hidden vector from GPT2. We believe adding this to a copy distribution from tokens in the previous questions could improve the performance $p_{copy}$. We believe this since the words used to reformulate the query often appear in the dialog history. The selection of the next query would then be $w p_{gen} + (1-w)p_{copy}$ where $w$ is a weight for adding the distributions. This is motivated by pointer-generator networks and their success with text summarization. \cite{see2017point}.

\section*{Acknowledgment}
The authors would to thanks Prof. Ali Ghodsi for sponsoring some Compute Canada resources for our project. Thanks to Prof. Charles Clarke and Negar Arabzadehghahyazi for pointing us to datasets.

\ifCLASSOPTIONcaptionsoff
  \newpage
\fi

\bibliographystyle{plain}

\end{document}